\documentclass{aastex}
\pdfoutput=1

\def\ltsima{$\; \buildrel < \over \sim \;$}
\def\simlt{\lower.5ex\hbox{\ltsima}} 
\def\gtsima{$\; \buildrel > \over \sim \;$}
\def\simgt{\lower.5ex\hbox{\gtsima}} 

\def\deg{$^\circ$}
\def\3eg{3EG~J093--3431}

\def\GeV{\hbox{~GeV}}
\def\MeV{\hbox{~MeV}}

\def\keV{\hbox{~keV}}

\shorttitle{VHE Constraints on GRB\,130427A } 
\shortauthors{VERITAS Collaboration}

\begin{document}

\title{Constraints on Very High Energy Emission from GRB\,130427A}

\author{
E.~Aliu\altaffilmark{1},
T.~Aune\altaffilmark{2,*},
A.~Barnacka\altaffilmark{3},
M.~Beilicke\altaffilmark{4},
W.~Benbow\altaffilmark{5},
K.~Berger\altaffilmark{6},
J.~Biteau\altaffilmark{7},
J.~H.~Buckley\altaffilmark{4},
V.~Bugaev\altaffilmark{4},
K.~Byrum\altaffilmark{8},
J.~V~Cardenzana\altaffilmark{9},
M.~Cerruti\altaffilmark{5},
X.~Chen\altaffilmark{10,11},
L.~Ciupik\altaffilmark{12},
V.~Connaughton\altaffilmark{13},
W.~Cui\altaffilmark{14},
H.~J.~Dickinson\altaffilmark{9},
J.~D.~Eisch\altaffilmark{9},
M.~Errando\altaffilmark{1},
A.~Falcone\altaffilmark{15},
S.~Federici\altaffilmark{10,11},
Q.~Feng\altaffilmark{14},
J.~P.~Finley\altaffilmark{14},
H.~Fleischhack\altaffilmark{11},
P.~Fortin\altaffilmark{5},
L.~Fortson\altaffilmark{16},
A.~Furniss\altaffilmark{7},
N.~Galante\altaffilmark{5},
G.~H.~Gillanders\altaffilmark{17},
S.~Griffin\altaffilmark{18},
S.~T.~Griffiths\altaffilmark{19},
J.~Grube\altaffilmark{12},
G.~Gyuk\altaffilmark{12},
N.~H{\aa}kansson\altaffilmark{10},
D.~Hanna\altaffilmark{18},
J.~Holder\altaffilmark{6},
G.~Hughes\altaffilmark{11},
T.~B.~Humensky\altaffilmark{20},
C.~A.~Johnson\altaffilmark{7},
P.~Kaaret\altaffilmark{19},
P.~Kar\altaffilmark{21},
M.~Kertzman\altaffilmark{22},
Y.~Khassen\altaffilmark{23},
D.~Kieda\altaffilmark{21},
H.~Krawczynski\altaffilmark{4},
F.~Krennrich\altaffilmark{9},
M.~J.~Lang\altaffilmark{17},
A.~S~Madhavan\altaffilmark{9},
G.~Maier\altaffilmark{11},
S.~McArthur\altaffilmark{24},
A.~McCann\altaffilmark{25},
K.~Meagher\altaffilmark{26},
J.~Millis\altaffilmark{27},
P.~Moriarty\altaffilmark{17,28},
R.~Mukherjee\altaffilmark{1},
D.~Nieto\altaffilmark{20},
A.~O'Faol\'{a}in de Bhr\'{o}ithe\altaffilmark{11},
R.~A.~Ong\altaffilmark{2},
A.~N.~Otte\altaffilmark{26},
N.~Park\altaffilmark{24},
M.~Pohl\altaffilmark{10,11},
A.~Popkow\altaffilmark{2},
H.~Prokoph\altaffilmark{11},
E.~Pueschel\altaffilmark{23},
J.~Quinn\altaffilmark{23},
K.~Ragan\altaffilmark{18},
J.~Rajotte\altaffilmark{18},
L.~C.~Reyes\altaffilmark{29},
P.~T.~Reynolds\altaffilmark{30},
G.~T.~Richards\altaffilmark{26},
E.~Roache\altaffilmark{5},
G.~H.~Sembroski\altaffilmark{14},
K.~Shahinyan\altaffilmark{16},
A.~W.~Smith\altaffilmark{21},
D.~Staszak\altaffilmark{18},
I.~Telezhinsky\altaffilmark{10,11},
J.~V.~Tucci\altaffilmark{14},
J.~Tyler\altaffilmark{18},
A.~Varlotta\altaffilmark{14},
V.~V.~Vassiliev\altaffilmark{2},
S.~Vincent\altaffilmark{11},
S.~P.~Wakely\altaffilmark{24},
O.~M.~Weiner\altaffilmark{20},
A.~Weinstein\altaffilmark{9},
R.~Welsing\altaffilmark{11},
A.~Wilhelm\altaffilmark{10,11},
D.~A.~Williams\altaffilmark{7},
B.~Zitzer\altaffilmark{8},
J.~E.~McEnery\altaffilmark{31,32}, 
J.~S.~Perkins\altaffilmark{31}, 
P.~Veres\altaffilmark{33,34,**}, 
S.~Zhu\altaffilmark{32,***}
}

\altaffiltext{1}{Department of Physics and Astronomy, Barnard College, Columbia University, NY 10027, USA}
\altaffiltext{2}{Department of Physics and Astronomy, University of California, Los Angeles, CA 90095, USA}
\altaffiltext{3}{Harvard-Smithsonian Center for Astrophysics, 60 Garden Street, Cambridge, MA 02138, USA}
\altaffiltext{4}{Department of Physics, Washington University, St. Louis, MO 63130, USA}
\altaffiltext{5}{Fred Lawrence Whipple Observatory, Harvard-Smithsonian Center for Astrophysics, Amado, AZ 85645, USA}
\altaffiltext{6}{Department of Physics and Astronomy and the Bartol Research Institute, University of Delaware, Newark, DE 19716, USA}
\altaffiltext{7}{Santa Cruz Institute for Particle Physics and Department of Physics, University of California, Santa Cruz, CA 95064, USA}
\altaffiltext{8}{Argonne National Laboratory, 9700 S. Cass Avenue, Argonne, IL 60439, USA}
\altaffiltext{9}{Department of Physics and Astronomy, Iowa State University, Ames, IA 50011, USA}
\altaffiltext{10}{Institute of Physics and Astronomy, University of Potsdam, 14476 Potsdam-Golm, Germany}
\altaffiltext{11}{DESY, Platanenallee 6, 15738 Zeuthen, Germany}
\altaffiltext{12}{Astronomy Department, Adler Planetarium and Astronomy Museum, Chicago, IL 60605, USA}
\altaffiltext{13}{Center for Space Plasma and Aeronomic Research (CSPAR), University of Alabama in Huntsville, Huntsville, AL 35899, USA}
\altaffiltext{14}{Department of Physics and Astronomy, Purdue University, West Lafayette, IN 47907, USA}
\altaffiltext{15}{Department of Astronomy and Astrophysics, 525 Davey Lab, Pennsylvania State University, University Park, PA 16802, USA}
\altaffiltext{16}{School of Physics and Astronomy, University of Minnesota, Minneapolis, MN 55455, USA}
\altaffiltext{17}{School of Physics, National University of Ireland Galway, University Road, Galway, Ireland}
\altaffiltext{18}{Physics Department, McGill University, Montreal, QC H3A 2T8, Canada}
\altaffiltext{19}{Department of Physics and Astronomy, University of Iowa, Van Allen Hall, Iowa City, IA 52242, USA}
\altaffiltext{20}{Physics Department, Columbia University, New York, NY 10027, USA}
\altaffiltext{21}{Department of Physics and Astronomy, University of Utah, Salt Lake City, UT 84112, USA}
\altaffiltext{22}{Department of Physics and Astronomy, DePauw University, Greencastle, IN 46135-0037, USA}
\altaffiltext{23}{School of Physics, University College Dublin, Belfield, Dublin 4, Ireland}
\altaffiltext{24}{Enrico Fermi Institute, University of Chicago, Chicago, IL 60637, USA}
\altaffiltext{25}{Kavli Institute for Cosmological Physics, University of Chicago, Chicago, IL 60637, USA}
\altaffiltext{26}{School of Physics and Center for Relativistic Astrophysics, Georgia Institute of Technology, 837 State Street NW, Atlanta, GA 30332-0430}
\altaffiltext{27}{Department of Physics, Anderson University, 1100 East 5th Street, Anderson, IN 46012}
\altaffiltext{28}{Department of Life and Physical Sciences, Galway-Mayo Institute of Technology, Dublin Road, Galway, Ireland}
\altaffiltext{29}{Physics Department, California Polytechnic State University, San Luis Obispo, CA 94307, USA}
\altaffiltext{30}{Department of Applied Physics and Instrumentation, Cork Institute of Technology, Bishopstown, Cork, Ireland}
\altaffiltext{31}{NASA Goddard Space Flight Center, Greenbelt, MD 20771, USA}
\altaffiltext{32}{Department of Physics and Department of Astronomy, University of Maryland, College Park, MD 20742, USA}
\altaffiltext{33}{Department of Physics, George Washington University, Washington, DC 20052, USA}
\altaffiltext{34}{Department of Astronomy and Astrophysics, Department of Physics, and Center for Particle and Gravitational Astrophysics, Pennsylvania State University, University Park, PA 16802, USA}

\altaffiltext{*}{Corresponding author: aune@astro.ucla.edu}
\altaffiltext{**}{Corresponding author: veres@email.gwu.edu}
\altaffiltext{***}{Corresponding author: sjzhu@umd.edu}

\slugcomment{Accepted by ApJ}

\begin{abstract}
Prompt emission from the very fluent and nearby ($z=0.34$)
gamma-ray burst GRB\,130427A was detected by several orbiting
telescopes and by ground-based, wide-field-of-view optical transient
monitors. Apart from the intensity and proximity of this GRB, it is
exceptional due to the extremely long-lived high-energy ($100$ MeV to
$100$ GeV) gamma-ray emission, which was detected by the Large Area
Telescope on the {\it Fermi Gamma-ray Space Telescope} for $\sim\!70$
ks after the initial burst. The persistent, hard-spectrum, high-energy
emission suggests that the highest-energy gamma rays may have been
produced via synchrotron self-Compton processes though there is also
evidence that the high-energy emission may instead be an extension of
the synchrotron spectrum.  VERITAS, a ground-based imaging atmospheric
Cherenkov telescope array, began follow-up observations of
GRB\,130427A $\sim\!71$ ks ($\sim\!20$ hr) after the onset of the
burst. The GRB was not detected with VERITAS; however, the high
elevation of the observations, coupled with the low redshift of the
GRB, make VERITAS a very sensitive probe of the emission from
GRB\,130427A for $E > 100$ GeV. The non-detection and consequent
upper limit derived place constraints on the synchrotron self-Compton
model of high-energy gamma-ray emission from this burst.
\end{abstract}
 

\section{Introduction\label{section-intro}}
Gamma-ray bursts (GRBs) are commonly thought to result from collapsing
massive stars or merging compact objects, which form a black hole or
neutron star. In the standard GRB model \citep[see, for
  example,][]{Piran:1999fk}, the initial bright prompt emission is
produced within a relativistic jet after it escapes through the
stellar envelope and
could produce radiation via a number of processes including internal
shocks, magnetic reconnection, or hydromagnetic turbulence. As the
ejecta sweep up external material, forward and reverse shocks are
created that can accelerate charged particles, producing MeV to GeV
gamma-ray photons via synchrotron radiation. It has been suggested
that GRBs might also create detectable fluxes of high-energy photons
at later times via synchrotron self-Compton or external Compton
processes \citep{Zhang:2001ws, Wang:2001kx, Beloborodov:2005uq,
  Wang:2006hy}. In addition, the external shocks themselves could
produce very high energy (VHE, $E>100$ GeV) photons via the inverse
Compton mechanism \citep{Meszaros:1994vn, Dermer:2000du,
  Fan:2008wg, Sari+01ic}. These non-thermal processes could produce photons with
energies as high as $\sim 1$ TeV in the early afterglow phase of the GRB.

The extraordinary GRB\,130427A was initially detected at 07:47:06.42
UTC \citep{vonKienlin:2013} on 2013 April 27 by the Gamma-ray Burst
Monitor \citep[GBM,][]{Meegan:2009ys} on board the {\it Fermi
  Gamma-ray Space Telescope}.  This detection triggered an autonomous
repoint request that kept the burst in the field of view (FoV) of the
Large Area Telescope \citep[LAT,][]{Atwood:2009pf} for 2.5 hr
except during periods of Earth occultation \citep{Ackermann:2013sc}.
The Burst Alert Telescope
\citep[BAT,][]{2005SSRv..120..143B} on board the {\it Swift}
observatory independently triggered on this burst at 07:47:57 UTC
\citep{Maselli:2014hc}.  The preliminary 15-350 keV BAT light curve
showed an extremely bright burst with a highly structured peak
lasting 20 s and displaying a maximum count rate of approximately
100,000 counts per second \citep{Maselli:2013dq}.

\citet{Levan:2013dq} determined that the GRB was associated with a
Type IC supernova (SN 2013cq) in a galaxy at a redshift of {\it z} =
0.34. The average redshift of {\it Swift}-detected long GRBs is $z>2$
\citep{Gehrels:2009p4268}. GRB\,130427A had the highest prompt fluence
yet recorded by Konus-{\it WIND} (20--1200 keV)
\citep{2013GCN..14487...1G} and {\it Fermi}-GBM
\citep{vonKienlin:2013}, as well as the longest-lasting emission and
highest observed photon energy (95 GeV) from a GRB yet recorded by
the {\it Fermi}-LAT \citep{Ackermann:2013sc}.  HAWC, a wide FoV, high
duty cycle, water Cherenkov detector currently under construction and
sensitive to gamma rays in the GeV--TeV energy range
\citep{Abeysekara:2013uq}, did not detect prompt VHE emission from
GRB\,130427A \citep{2013GCN..14549...1L}.

It has long been predicted that GRBs could emit gamma rays at energies
above 100 GeV. GRB\,080916C \citep{2013ApJ...774...76A} and
GRB\,130427A both produced photons with energies above 100 GeV, but
the detected photon energies were lower due to the cosmological
redshift of the bursts. No direct detection of $>100$ GeV photons has
yet been achieved even though significant effort has been put into
searching for such emission \citep{Connaughton:1997cg,
  Aharonian:2009iza, Albert:2007p1679, Acciari:2011hr, Atkins05}.

\begin{figure}
\epsscale{0.7}
\plotone{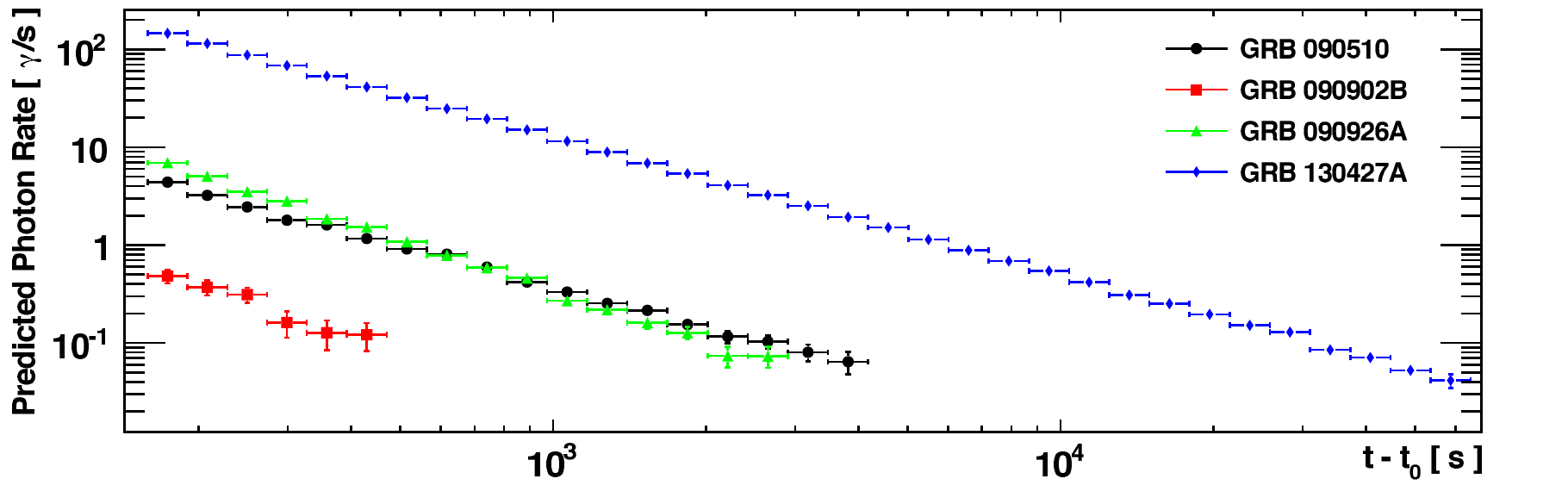}
\epsscale{0.75}
\plotone{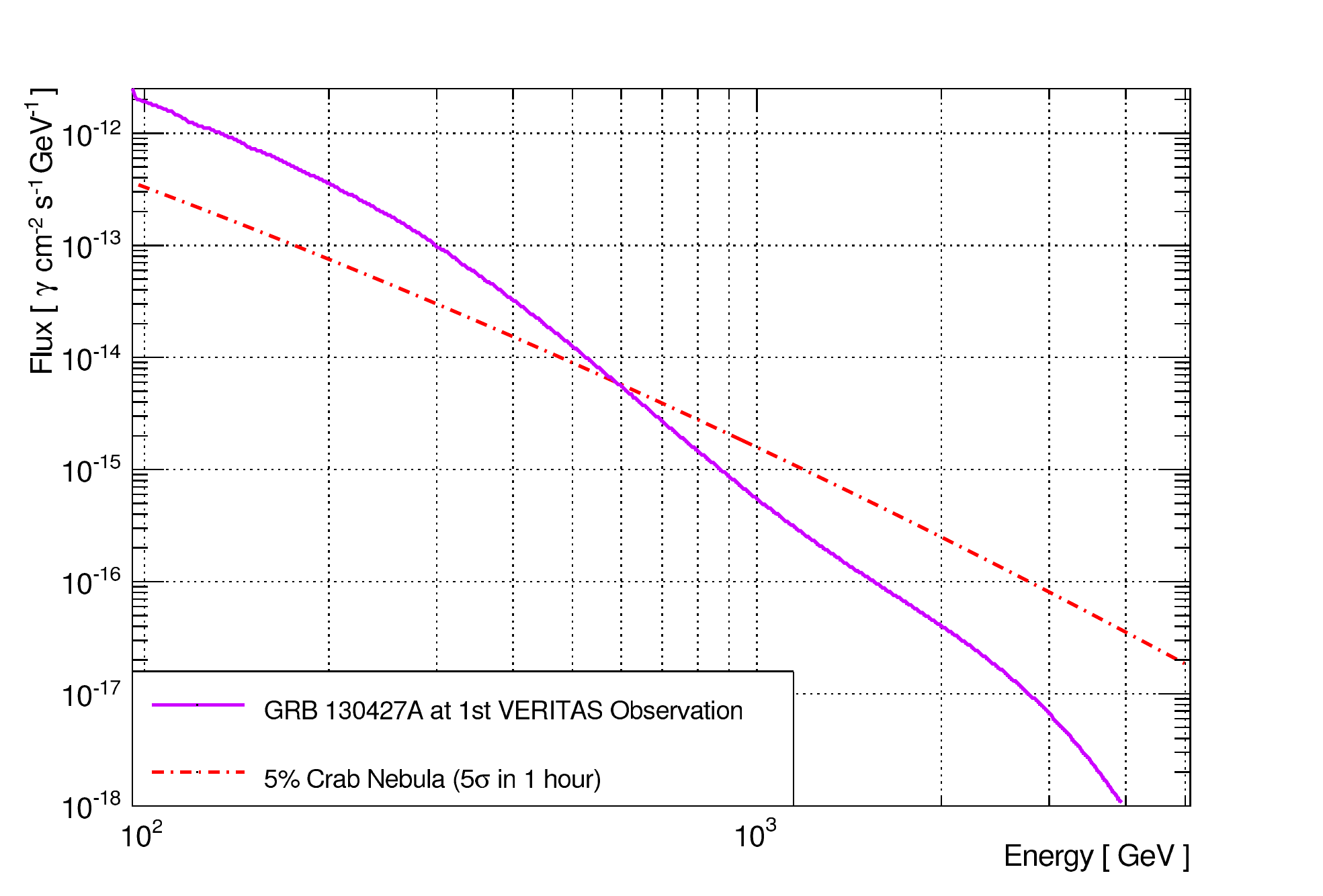}
\caption{\label{fig:prediction}\footnotesize{The upper panel shows
    predicted light curves for several fluent, LAT-detected GRBs:
    GRB\,090510 \citep{DePasquale:2010p2521}, GRB\,090902B
    \citep{2009ApJ...706L.138A}, GRB\,090926A
    \citep{Ackermann:2011p4979}, and GRB\,130427A
    \citep{Ackermann:2013sc}, as they would be seen by VERITAS at
    energies greater than 100 GeV assuming an elevation of 70
    degrees. Each bin in the upper panel is derived from a figure
    similar to that in the lower panel, which is a detail of the
    photon flux extrapolated from LAT data of GRB\,130427A (including
    EBL absorption) over the period of the first VERITAS observation
    (see Table \ref{tab:VTSObsTable}).  The red dashed-dotted line is
    5\% of the average Crab Nebula spectrum for reference. VERITAS is
    capable of detecting a 5\% Crab Nebula source over the duration of
    the first GRB\,130427A observation ($\sim 1$ hour). The lack of a
    detection by VERITAS suggests the presence of a spectral and/or
    temporal cutoff at high energies and late times, respectively.}}
\end{figure}

A simple extrapolation of the late-time LAT light curve to very high
energies, taking into account extragalactic background light
\citep[EBL,][]{Gould:1967bj} attenuation, indicates that
current-generation IACT arrays were sensitive enough to detect
GRB\,130427A up to about a day after the onset of the burst. The top
panel of Figure \ref{fig:prediction} shows the predicted light curves
for several bright LAT-detected GRBs as they would appear to VERITAS,
an IACT array sensitive to gamma rays above 100 GeV and located in
southern Arizona. All bins represent a detection of more than three
standard deviations above background ($>3 \sigma$). The predictions
use the fluxes and spectra from the LAT measurements, specifically
$dN/dt \propto t^{-1.35}$ and $dN/dE \propto E^{-2.2}$
\citep{Ackermann:2013sc} and include the absorption of gamma rays by
the EBL according to the model of \citet{Gilmore:2009p1285}.
GRB 130427A, shown in blue, is by
far the most promising candidate for a VHE detection by
VERITAS. VERITAS made observations toward the direction of the GRB
starting $\sim 20$ hr after the initial satellite detection but did
not detect any emission from the burst. This Letter details those
observations and places them in context with observations at other
wavelengths, especially those made by the LAT.  Additionally,
constraints on the VHE emission obtained from this non-detection are
discussed in the context of various emission models.

\begin{figure}
\plotone{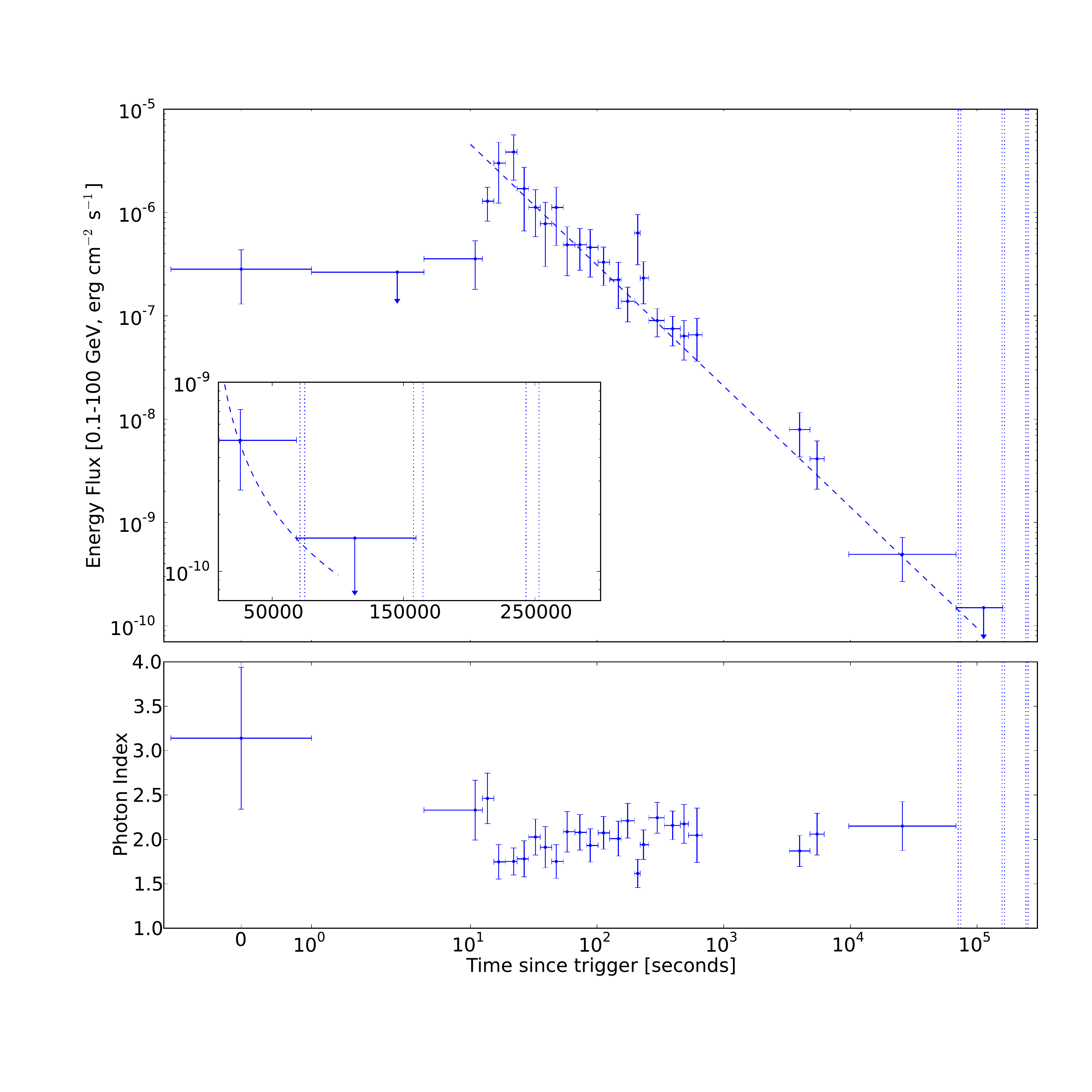}
\caption{\label{fig:lightcurve} Upper panel shows the 0.1--100
  GeV light curve for GRB\,130427A as measured by the LAT.  The dashed
  line is a power-law fit to the light curve.  The lower panel shows
  the LAT-measured photon index.  These data have been shown
  previously in \citet{Ackermann:2013sc}.  The vertical dotted lines
  indicate the times of the three VERITAS observations given in Table
  \ref{tab:VTSObsTable}.  The inset details these observations.}
\end{figure}

\section{Observations}
\subsection{VERITAS}

The energy range of the VERITAS array extends from $\sim\!100$ GeV
to several tens of TeV, overlapping with the energy range of the LAT
\citep[for an overview of VERITAS, see][]{Holder:2011wu}.  The VERITAS
Collaboration has had a GRB observing program since it began full
array operations in 2007 and has performed more than 100 follow-up
observations of GRBs detected by space-based instruments
\citep{Acciari:2011hr}. The VERITAS trigger system was upgraded in
2011 and the camera was upgraded one year later, resulting in improved
sensitivity and a lower energy threshold \citep{Kieda:2014}.  
It is estimated that VERITAS should be sensitive enough to detect bright
and/or nearby GRBs.

At the time of the {\it Fermi}-GBM trigger (07:47 UTC), GRB\,130427A
was at a relatively favorable elevation of 52\deg~for
VERITAS. Unfortunately, bright moonlight conditions (97\% full and
$\sim\!30$\deg~above the horizon) precluded observations. Typical GRB
follow-up observations are limited to three hours after a burst, but due
to the extraordinary nature of GRB\,130427A, VERITAS observations were
initiated the following night, 2013 April 28, at 03:32:35 (UTC),
71.128 ks after the {\it Fermi}-GBM trigger. Observations lasted for
59 minutes until moonrise.  Observations continued on the following
two nights, lasting $\sim\!2$ and $\sim\!2.5$ hours, respectively (see
Table \ref{tab:VTSObsTable}).  The average elevation of the GRB
position at the time of the observations was $81^{\circ}$, resulting
in a post-analysis energy threshold of $\sim\!100$ GeV.

\begin{table*}
  \begin{small}
  \caption{VERITAS Observations of GRB\,130427A}
  \label{tab:VTSObsTable}
  \begin{centering}
  \begin{tabular*}{\textwidth}{@{\extracolsep{\fill}}cccc|ccccc}
    \hline\hline
    Date & $t_{\rm start}$ & $t_{\rm end}$ & Exposure & 
    $n_{\rm on}$ & $n_{\rm off}$ & $\alpha^*$ & Significance$^\dagger$ & Flux UL$^{\ddagger}$ \\
     & (UTC) & (UTC) & (s) & & & & ($\sigma$) & \\
    \hline
    2013 April 28 & 03:32:35 & 04:31:16 & 2925 & 165 & 1164 & 0.125 & 1.3 & $9.4 \times 10^{-12}$ \\
    2013 April 29 & 03:32:59 & 05:33:39 & 5746 & 322 & 2120 & 0.143 & 1.1 & $6.6 \times 10^{-12}$ \\
    2013 April 30 & 03:22:02 & 06:05:40 & 7814 & 402 & 2820 & 0.147 & -0.5 & $2.7 \times 10^{-12}$ \\
    \hline
    Total & & & 16485 & 889 & 6104 & 0.141 & 0.9 & $3.3 \times 10^{-12}$ \\ 
    \hline\hline
  \end{tabular*}
\end{centering}
\end{small}
$^*$ Ratio of the signal region to background region. \\
$^\dagger$ Significance calculated using eq. 17 of \citet{Li:1983p575} \\
$^{\ddagger}$ 99\% confidence-level upper limit on $\nu \rm{F}_{\nu}$ in
  erg cm$^{-2}$ s$^{-1}$. The upper limit is derived using the method of 
  \citet{Rolke:2005p655}, quoted at 100 GeV, and calculated assuming an
  intrinsic GRB spectrum of $\frac{{\rm d}N} {{\rm d}E} \propto
  E^{-2.0}$ (as measured by the LAT) absorbed using the EBL model
  of \citet{Gilmore:2009p1285}.
\end{table*}

\subsection{{\it Fermi}-LAT}

The LAT is a pair-conversion telescope that detects photons with
energies from 20~MeV to $>$300~GeV \citep{Atwood:2009pf}. The GRB was
within the LAT FoV (47$^\circ$\kern-4pt.3 from the boresight) at the time of
the trigger and remained in the FoV for the next 2.5 hr due to the
autonomous repoint request (except during times of Earth occultation).
Once the observatory returned to survey mode, the GRB was in the FoV
$\sim$40\% of the time.  During the first VERITAS observation (71.0
to 75.0~ks), the GRB was in the LAT FoV from 72.1 to 73.4~ks and
73.5 to 74.9~ks; the last photon with energy greater than 1 GeV was
detected at 68.4~ks.

\section{Analysis \& Results}
\subsection{VERITAS}

The VERITAS data were analyzed with a standard VERITAS software
package using event selection criteria optimized for a soft-spectrum
($\frac{dN}{dE} \propto E^{-3.5}$), weak (5\% Crab Nebula flux) point
source, which roughly approximates the EBL-absorbed GRB spectrum.  We
decided, {\it a priori} to analyze the data from each night's
observations independently in addition to the complete data set
together. We find no evidence for gamma-ray emission above 100 GeV in
any analysis. This result is confirmed by an analysis using an
independent software package.

We derive upper limits on the VHE gamma-ray flux from
GRB\,130427A. The assumed spectral shape is extrapolated from the LAT
observations, namely a power-law spectrum with a photon index of
$2$ with no intrinsic cutoff. The upper limits calculated for
each time interval are given in Table \ref{tab:VTSObsTable}.

\subsection{{\it Fermi}-LAT}

We analyzed the LAT data using an unbinned maximum likelihood method
(as implemented in the Fermi Science Tools
v9r30p1\footnote{http://fermi.gsfc.nasa.gov/ssc/data/analysis/software/}). The
spectrum of the GRB is modeled as a power law and the background is
modeled using the standard Galactic and isotropic diffuse models,
specifically gal\_2yearp7v6\_v0.fits and
iso\_p7v6source.txt\footnote{http://fermi.gsfc.nasa.gov/ssc/data/access/lat/BackgroundModels.html};
there were no LAT point sources in the region bright enough to warrant
inclusion in the source model.  Pass 7 {\it Source} class events
within a $10^\circ$ region around the burst position \citep[R.A. =
  $11^{\rm h} 32^{\rm m} 32.82^{\rm s}$ Dec. = $+27^{\circ}
  41'56.06''$, J2000,][]{Perley:2013} were used with the standard
zenith angle cut of $100^\circ$ (to limit contamination from the
gamma-ray bright limb of the Earth) and the appropriate set of
instrument response functions (P7\_SOURCE\_V6). The LAT emission
decays smoothly after the first $\sim$20 s; the energy flux
light curve is well fitted with a single power law with a temporal
index of $-1.17 \pm 0.06$, and the photon flux light curve is well fitted
by a broken power law with a temporal index of $-0.85 \pm 0.08$ before
$t-t_0 = 381 \pm 138$ s and a temporal index of $-1.35 \pm 0.08$ at
later times (see \citep{Ackermann:2013sc} for details of the
analysis).

We considered the LAT emission between 10~ks and 70~ks after the
burst. This was the last time interval before the VERITAS observations
during which the LAT detected significant flux (Figure
\ref{fig:lightcurve}), as determined in \citep{Ackermann:2013sc}.
To test for spectral curvature, we also fitted the
data with a power law with an exponential cutoff and with a broken
power law. Neither of these models is statistically preferred over
the simple power law. The spectrum of the GRB in this time interval is
consistent with its spectrum earlier during the burst (the photon
index is $2.2 \pm 0.2$, see Figure \ref{fig:lightcurve}). The LAT
data are best fitted with a power-law $\frac{dN} {dE} = N_0(E/E_0$)$^{-\gamma}$
where $E_0$, the decorrelation energy, is 826 MeV, $\gamma$ is $2.2 \pm
0.2$, and $N_0$ is $6.7 \pm 2.0 \times 10^{-11}$ cm$^{-2}$ s$^{-1}$ MeV$^{-1}$.
This decorrelation energy is uniquely determined by the fit of the spectral 
index and integral flux over the energy range of the LAT and is the 
energy at which the normalization and spectral index are the least correlated.

\section{Discussion \& Conclusions}

\begin{figure}
\epsscale{1}
\plotone{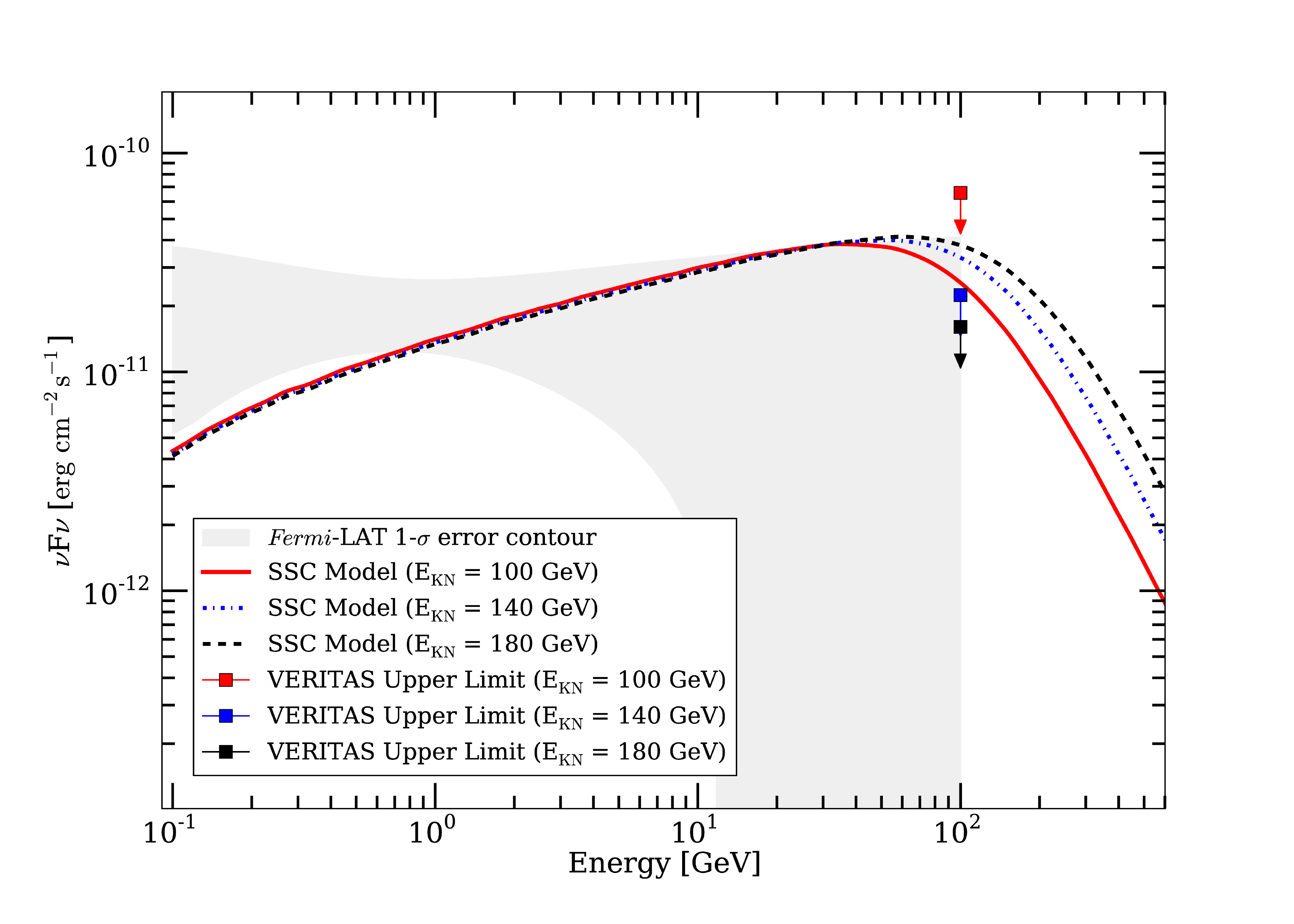}
\caption{\label{fig:sed} Joint VERITAS-LAT spectral energy
  distribution. The VERITAS upper limits are calculated assuming an
  SSC model \citep{Sari+01ic} with an electron spectrum $\frac{dN}
  {dE} \propto E^{-2.45}$ and breaks at 100, 140, and 180 GeV (solid,
  dot-dashed, and dashed lines). The electron energy distribution is
  determined from the LAT-measured spectrum, as described in the
  text. This SED is then absorbed using the EBL model of
  \citet{Gilmore:2009p1285}. The LAT data are best fitted with a power
  law with an index of $2.2 \pm 0.2$.  The gray shaded region (the
  ``bowtie'') shows the one sigma range of power-law models compatible
  with the LAT data after extrapolating from the last LAT time bin
  (10~ks to 70~ks) into the VERITAS observing time (71~ks to 75~ks)
  using the photon flux relation $ \frac{dN}{dt} \propto t^{-1.35 \pm
    0.08}$, which was obtained from fitting the late-time LAT data
  \citep{Ackermann:2013sc}. The electron spectral index of the SSC
  models is determined from the error-weighted mean of the late-time
  spectral and temporal indices measured by the LAT.}
\end{figure}

The VERITAS upper limit and the last significant detection of
high-energy emission by the LAT are not simultaneous. However, the
late-time emission ($>\!200$ s) measured by the LAT shows no deviation
from a well-defined power-law behavior in both time and energy (see
Figure \ref{fig:lightcurve}), so we extrapolate the LAT data to the
first VERITAS observing interval using the photon flux relation $dN/dt
\propto t^{-1.35 \pm 0.08}$ measured by the LAT to create the joint
VERITAS-LAT spectral energy distribution (SED) shown in Figure \ref{fig:sed}. While compatible with
the extrapolation of the LAT measurement, the VERITAS upper limits
disfavor a scenario in which there is an enhanced VHE component. Both
synchrotron \citep[e.g.][]{Kouv+13nustar} and inverse Compton
\citep[e.g.][]{Liu:2013vg} scenarios have been proposed to explain the
late-time, high-energy emission from GRB\,130427A and we briefly
examine these models in the context of the VERITAS upper limit.

\citet{Ackermann:2013sc} noted that the synchrotron interpretation is
problematic for this burst due to the observed late-time, high-energy
photons, which contradict the robust limits obtained from a simple
interpretation of the radiation produced in shocked plasma. However,
\citet{Kouv+13nustar} find that both spectral and temporal
extrapolations, from optical to multi-GeV energies, are consistent
with the synchrotron mechanism, though such an interpretation requires
significant modifications to current models of particle acceleration
in GRB afterglow shocks. In the context of the synchrotron model, we
interpret the VERITAS upper limit in a scenario where the uniform
magnetic field assumption in the shocked interstellar medium (ISM) is
relaxed \citep{Kumar+12maxsyn}, and the magnetic field decays as a
power law in the shocked region. Bearing in mind the assumptions of
this model, the VERITAS non-detection can be associated with a cutoff
in the synchrotron photon spectrum at $\sim\!100 \GeV$.  The
theoretical limit on the synchrotron cutoff energy can be expressed as
\[ E_{\rm{cut,syn}} = 50 \MeV \left [ \frac{\Gamma}{1+z} \right ] (B_w/B_0)\] 
\citep{Kumar+12maxsyn}. Here, $B_w$ is the magnetic field immediately
behind the shock front and it carries a fraction ($\epsilon_B$) of the
shocked gas energy density. $B_0$ is the shock-compressed magnetic
field of the ISM behind the thin shell associated with
the shock itself ($B_0\approx 4 \Gamma B_{\rm ISM}$).  The Lorentz
factor of the relativistic blast is derived from the self-similar
phase of the Blandford-McKee model \citep{Blandford:1976bo} and can be
written as $\Gamma \approx 10 (E_{53}/n_0)^{1/8} (t/t_{V})^{-3/8}$,
where $E_{53}$ is the isotropic equivalent energy of the burst in
units of $10^{53}$ ergs, $n_0=1$ cm$^{-3}$ is the density of the ISM,
and $t_{V}$ is the time of the first VERITAS observation. We limit our
discussion here to the ISM environment \citep{Maselli:2014hc}, but we
note that a wind \citep[e.g.][]{Perley:2014gk} or hybrid environment
\citep[$n \propto R^{-1.4\,\pm\,0.2}$, ][]{Kouv+13nustar} may instead
reflect the conditions surrounding GRB\,130427A. The VERITAS upper
limit can thus constrain the $B_w/B_0$ ratio to be $\gtrsim\!200$
where $B_w = \sqrt{32 \pi m_p \epsilon_B n_0} \Gamma c$
\citep{SariPiranNarayan98}. This upper limit can in turn be used to
constrain the magnetic field of the ISM in the GRB environment, which
can be written as $B_{\rm ISM} \lesssim 5 {\rm \mu G}\,E_{53}^{1/8}
\epsilon_{B,-4}^{1/2} n_0^{3/8} (t/t_V)^{-3/8}$ where $\epsilon_{B,-4}
= 10^{-4} \epsilon_B$.

It is also possible that the late-time, high-energy emission detected
by the LAT was produced from inverse-Compton scattering in a

high-energy emission in an SSC model is synchrotron photons
upscattered by the same electron population from which they were
emitted \citep{Sari+01ic}. For reasonable parameters, it can be shown
that the $0.1\lesssim E \lesssim 100 \GeV$ energy range falls between
the characteristic ($E^{\rm SSC}_m\approx 2\gamma_m^2 E_m \approx 25
\keV ~(t/t_V)^{-9/4}$) and cooling ($E^{\rm SSC}_c\approx 2\gamma_c^2
E_c\approx 400 \GeV ~(t/t_V)^{1/4}$) SSC energies ($\gamma_m$ and
$\gamma_c$ are the electron Lorentz factors at the minimum injection
energy and the cooling energy respectively). At late times, the LAT
measures a photon index of $2.2 \pm 0.2$ and a temporal index of
$-1.35 \pm 0.08$. Under the SSC model, both quantities can be used to
obtain the momentum power-law index for the shock-accelerated
electrons. For the input to the model, we use the error-weighted mean
of the electron power-law indices determined by each method to obtain
a shocked electron power-law index of $p=2.45$. It should be
noted that though this choice of electron index is appropriate given
the data, the spectral and temporal flux indices obtained from the SSC
model with this assumption are only consistent with the LAT
measurements at the level of two standard deviations. The
Klein-Nishina energy is $E_{\rm KN}=\Gamma \gamma_c m_e c^2/(1+z)
\approx 180 \GeV~(t/t_V)^{1/4}$ \citep{Guetta+03plerion}, where
$\Gamma\approx 10 (t/t_V)^{-3/8}$ is the bulk Lorentz factor of the
forward shock. Above this energy, the electron-photon scattering cross
section is reduced, resulting in a softening of the spectrum.

Figure \ref{fig:sed} shows the expected flux from SSC models fitted to
the late-time ($t-t_0 > 10$ ks) LAT-detected emission and with breaks
at 100, 140, and 180 GeV. The SSC models used here are taken from the
slow-cooling scenario described in \citet{Sari+01ic}. Also plotted is
the one sigma range of power-law models compatible with the LAT data
from the last LAT time bin ($10\, \mathrm{ks} < t-t_0 < 70\,
\mathrm{ks}$) extrapolated to the VERITAS observation time, as well as
the VERITAS upper limits for the three spectral assumptions. The
VERITAS upper limits are incompatible with a spectral break above
$\sim\!120$ GeV, or the absence of a break entirely. When the SSC
model spectrum, which is determined from the temporally extrapolated
LAT data, is extrapolated to energies above $\sim\!100$ GeV in the
model, the predicted flux conflicts with the upper limits obtained
with VERITAS. This indicates that the simple single zone SSC model is
not an accurate description of GRB\,130427A at energies greater than
$\sim\!100$~GeV.  It should be noted that another possible explanation
for the break is by a pair production cutoff of $100 \GeV$ photons
with $\sim\!100 \keV$ photons; however we find the optical depth for
this process is very low $\tau_{\gamma\gamma}\sim 10^{-5}$.  Thus we
conclude that the most plausible interpretation in the framework of an
inverse-Compton scenario is that we are observing the Klein-Nishina
cutoff below the VHE range. Results presented in this work, combined
with observations of GRB\,130427A at lower energies, suggest a single
dominant component in the afterglow. In order for this SSC
interpretation to work, we need a fine tuning of the underlying
physical parameters to: a) have an SSC-dominated afterglow from the
earliest times, or b) transition smoothly from synchrotron to
SSC-dominated regimes at late times. For this reason we prefer the
synchrotron interpretation of GRB\,130427A.

The VERITAS observations of GRB\,130427A, even at $\sim\!20$ hr after
the burst, meaningfully constrain synchrotron and inverse Compton
emission models that seek to explain the late-time, high-energy
emission observed by the LAT. Although it is estimated that a burst as
nearby as GRB\,130427A will occur only once every several decades, it
has been shown that bright bursts even out to $z \approx 2$ could be
detectable by VERITAS \citep{Acciari:2011hr}.  VERITAS continues to
perform follow-up observations of satellite-detected GRBs and efforts
to improve these observations are currently underway \citep{daw}.

\section{Acknowledgments}

VERITAS is supported by grants from the U.S. Department of Energy
Office of Science, the U.S. National Science Foundation and the
Smithsonian Institution, by NSERC in Canada, by Science Foundation
Ireland (SFI 10/RFP/AST2748) and by STFC in the U.K.  Additional
support for observations of GRBs comes from NASA grant NNX12AE30G. We
acknowledge the excellent work of the technical support staff at the
Fred Lawrence Whipple Observatory and at the collaborating
institutions in the construction and operation of the instrument.

The Fermi LAT Collaboration acknowledges generous ongoing
support from a number of agencies and institutes that have supported
both the development and the operation of the LAT as well as
scientific data analysis.  These include the National Aeronautics and
Space Administration and the Department of Energy in the United
States, the Commissariat \`a l'energie atomique and the Centre
national de la recherche scientifique / Institut national de physique
nucl\'eaire et de physique des particules in France, the Agenzia
Spaziale Italiana and the Istituto Nazionale di Fisica Nucleare in
Italy, the Ministry of Education, Culture, Sports, Science and
Technology (MEXT), High Energy Accelerator Research Organization (KEK)
and Japan Aerospace Exploration Agency (JAXA) in Japan, and the
K.~A.~Wallenberg Foundation, the Swedish Research Council and the
Swedish National Space Board in Sweden.

Additional support for science analysis during the operations phase is
gratefully acknowledged from the Istituto Nazionale di Astrofisica in
Italy and the Centre National d'\'Etudes Spatiales in France.

\end{document}